\documentclass[a4paper,11pt]{article}
\usepackage{ILD}
\usepackage[symbol]{footmisc}
\usepackage{feynmf}
\usepackage{array,multirow,graphicx}
\usepackage{graphicx}
\usepackage{cleveref}
\usepackage{tikz}
\usepackage[compat=1.1.0]{tikz-feynman}
\usetikzlibrary{arrows,shapes,positioning}
\usetikzlibrary{decorations.markings}
\usetikzlibrary{decorations.pathmorphing}
\usetikzlibrary{decorations.pathreplacing}
\usetikzlibrary{patterns}
\usetikzlibrary{plotmarks}
\usetikzlibrary{shadows}
\usepackage{contour}
\usepackage[style=numeric-comp,sorting=none]{biblatex}

\usepackage{subcaption}
\usepackage{wrapfig}

\title{Kinematic Fitting for ParticleFlow Detectors at Future Higgs Factories}

\ildproc{phys}{2021}{010}

\date{\today}

\addauthor{Yasser Radkhorrami}{\institute{1}\institute{2}}
\addauthor{Jenny List}{\institute{1}}

\addinstitute{1}{Deutsches Elektronen-Synchrotron (DESY), Hamburg}
\addinstitute{2}{Universit\"at Hamburg, Hamburg}

\abstract{In many analyses in Higgs, top and electroweak physics, the kinematic reconstruction of the final state is improved by constrained fits. This is a particularly powerful tool at $e^{+}e^{-}$ colliders, where the initial state four-momentum is known and can be employed to constrain the final state. A crucial ingredient to kinematic fitting is an accurate estimate of the measurement uncertainties, in particular for composed objects like jets. This contribution will show how the particle flow concept, which is a design-driver for most detectors proposed for future Higgs factories, can --- in addition to an excellent jet energy measurement --- provide detailed estimates of the covariance matrices for each individual particle-flow object (PFO) and each individual jet. Combined with information about leptons and secondary vertices in the jets, the kinematic fit enables to correct $b$- and $c$-jets for missing momentum from neutrinos from semi-leptonic heavy quark decays. The impact on the reconstruction of invariant di-jet masses and the resulting improvement in $ZH$ vs $ZZ$ separation will be presented, using the full simulation of the ILD detector, as an example of highly-granular ParticleFlow optimized detector concept.}

\titlecomment{Based on the talk "Kinematic Fitting for ParticleFlow Detectors at Future Higgs Factories" presented at The European Physical Society Conference on High Energy Physics (EPS-HEP2021), Hamburg, Germany, 26-30 July 2021, T12-400}

\graphicspath{ {./logos/}{./figures/} }

\begin{document}

\titlepage

\section{Introduction}\label{SEC:Intro}
\begin{figure}[htbp] \centering
	\begin{subfigure}{0.49\textwidth}
		\includegraphics[width=0.99\textwidth]{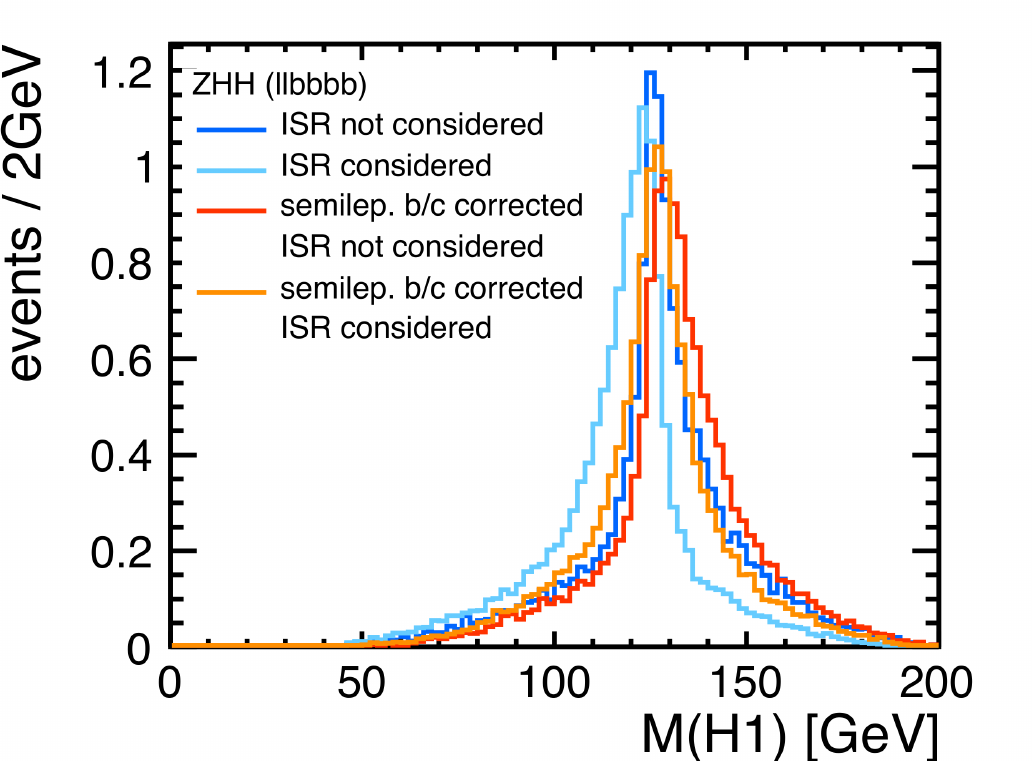}
		\label{fig:ZHH_Higgsmass}
	\end{subfigure}
	\begin{subfigure}{0.49\textwidth}
		\includegraphics[width=0.99\textwidth]{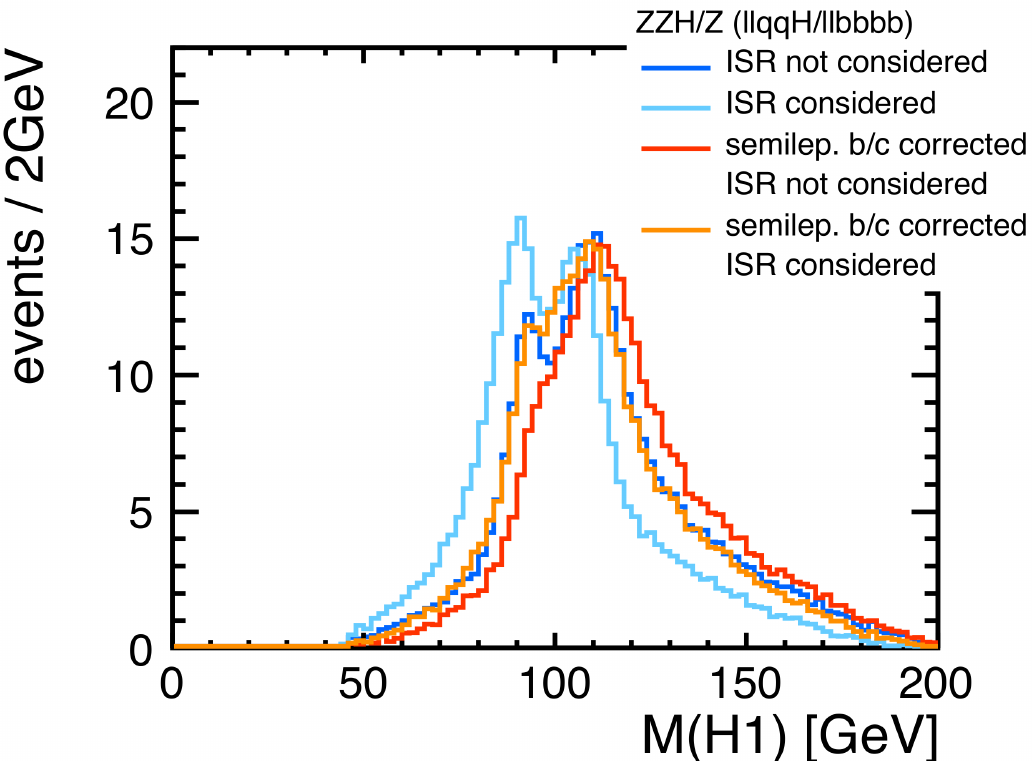}
		\label{fig:ZZHZ_HiggsmassBG}
	\end{subfigure}
	\caption{Higgs mass reconstruction in the presence of ISR photon and semi-leptonic decays. (a): the signal peak gets sharper by considering ISR photon and correction of semi-leptonic decays. (b): the background peak is pulled towards the signal.}
	\label{fig:ZZHZ_Higgsmass}
\end{figure}

The Higgs boson decay modes to heavy $b$ and $c$ quarks are crucial for the Higgs physics studies. Undetectable neutrinos due to semi-leptonic decays (SLD) in the jets originating from heavy $b$ and $c$ quarks degrade the reconstruction of $b$- and $c$-jets \cite[fig 8.3d]{theildcollaboration2020international}. A kinematic fit is a mathematical approach that can be used to retrieve the jet energy resolution beyond the detector resolution \cite{List:88030}. A key input to a kinematic fit is the measurement uncertainties. The ILD detector is based on the reconstruction of individual particles and provides an unprecedented knowledge about the jet-level uncertainties \cite{theildcollaboration2020international}. Already a very simple approximation of the energy of a missing neutrino in addition to parametrization of the jet energy error improves the di-jets invariant mass in $e^{+}e^{-}\rightarrow ZHH$ events at $\sqrt{s}=$ 500 GeV with $H\rightarrow b\bar{b}$, shown in fig \ref{fig:ZZHZ_Higgsmass} \cite{Duerig:310520}.
Despite the improvements in the Higgs mass reconstruction, the method applied in \cite{Duerig:310520} does not have a satisfying outcome in the presence of $ZZH/Z$ background. Thus a proper neutrino correction and a better jet error parametrization are needed to avoid the background being pulled towards the signal due to too much flexibility in the fit. This new jet error parametrization is evaluated using the $ZH/ZZ\rightarrow\mu\bar{\mu}b\bar{b}$ samples in presence of ISR photon and $\gamma\gamma\rightarrow$ low $p_{T}$ hadrons (only for $ZZ$ events) at $\sqrt{s}$=250 GeV.
\section{Kinematic fitting}\label{SEC:KinFit}
\begin{figure}[htbp] \centering
	\begin{subfigure}{0.48\textwidth}
		\includegraphics[width=0.99\textwidth]{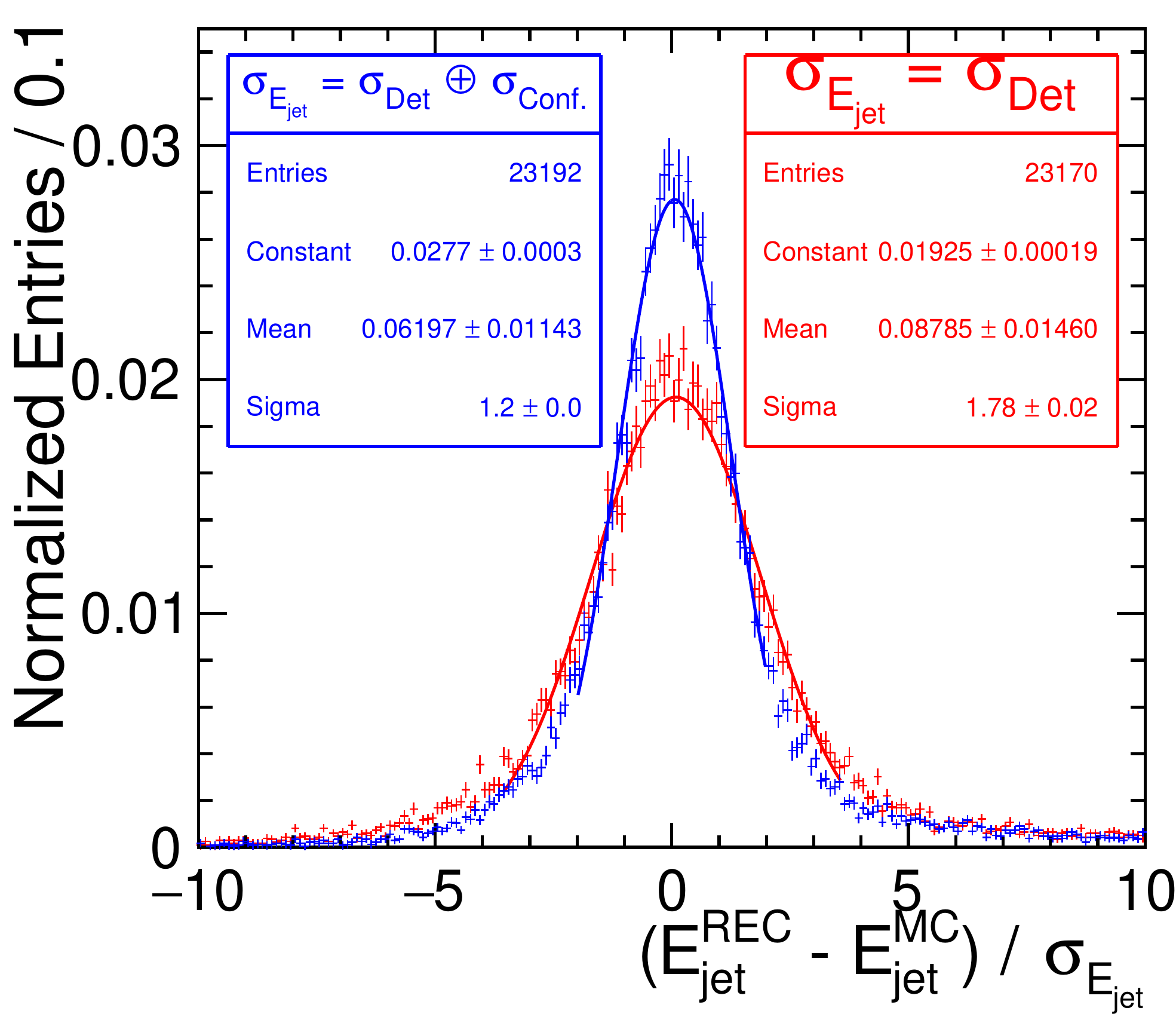}
		\label{fig:sigmaE_jet}
	\end{subfigure}
	\begin{subfigure}{0.51\textwidth}
		\includegraphics[width=0.93\textwidth]{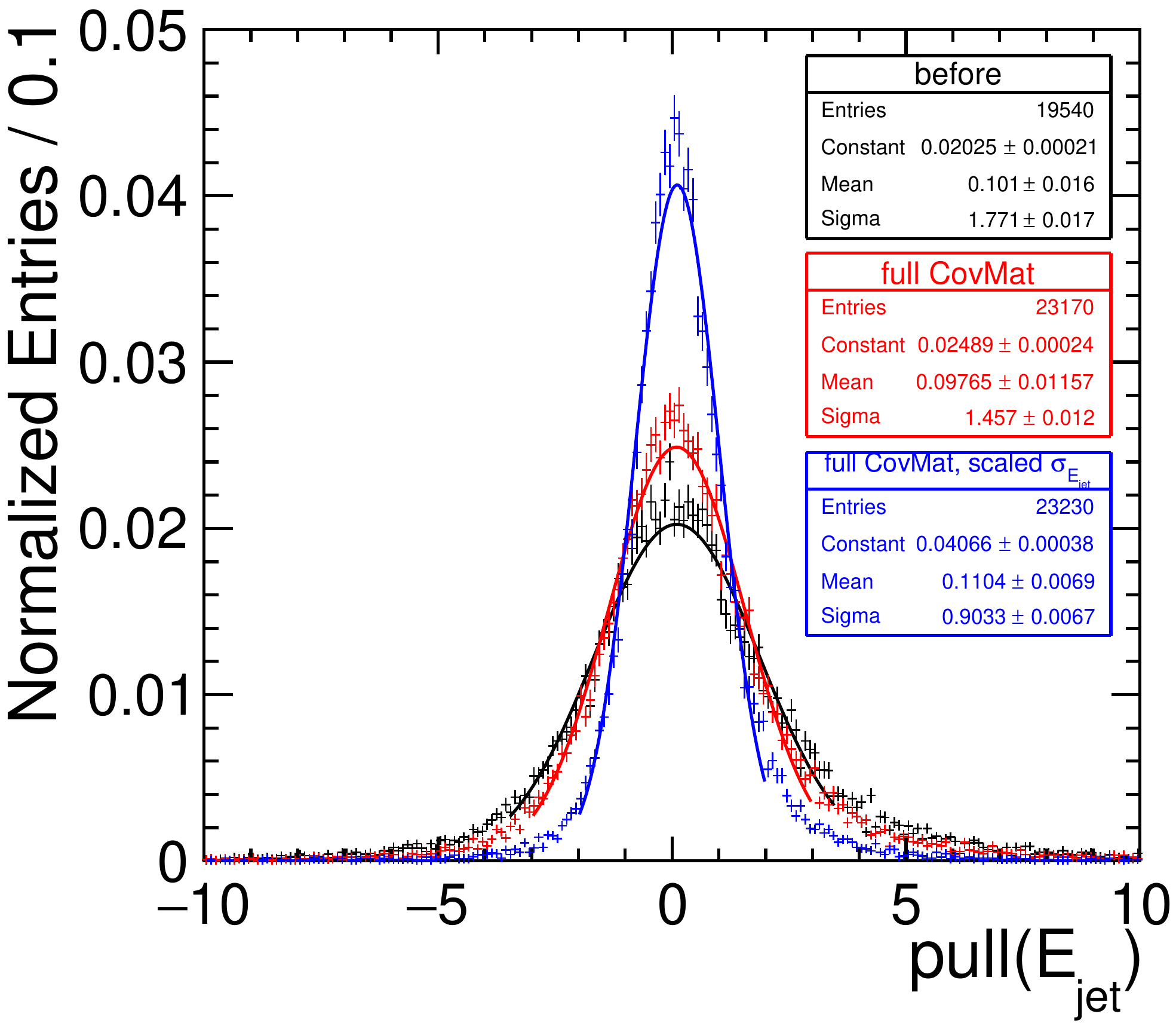}
		\label{fig:Pull_jetEnergy}
	\end{subfigure}
	\caption{(a): normalised residual of the jet energy when using only the sum of the PFO covariances (red) and when adding the confusion term derived in \cite{Ebrahimi:2017} (blue). (b): pull of the jet energies in the kinematic fit when using the previous error parametrization (black),  the parametrization corresponding to the blue histogram in fig \ref{fig:sigmaE}a (red) and when scaling $\sigma_{E}$ up by a factor 1.2 (blue), as motivated by the width of the normalised residual in fig \ref{fig:sigmaE}a.}
	\label{fig:sigmaE}
\end{figure}

It can be shown that the energy and the momentum of the missing neutrino can be obtained using the kinematics of the semi-leptonic decay \cite{radkhorrami2021conceptual}. In this method, the semi-leptonic decays are found by tagging charged leptons within $b$- or $c$-jets. The energy and momentum of the neutrino is then obtained up to a sign ambiguity if the four-momentum of visible decay products of the semi-leptonic decay, mass and flight direction of the parent hadron are known. The sign ambiguity is resolved by a kinematic fit \cite{List:88030}, imposing constraints on energy and momentum conservation and/or invariant masses of known particles. The Pandora particle flow algorithm \cite{Thomson_2009} used for event reconstruction in the ILD detector concept, provides full details of the measurement uncertainties for each PFO. Each of the ILD sub-detectors has their own spatial and energy resolution that affect the measurement error of particle reconstruction. For complicated cases like jets, the initial covariance matrix ($\sigma_{det}$) is formed by summing up the covariance matrices of all particles that belong to the jet (red histogram in fig \ref{fig:sigmaE}a).
Any confusion between hits created by charged and neutral PFOs will induce another error to the jet covariance matrix. A parametrization of the uncertainty due to confusion ($\sigma_{conf}$) has been derived for different jet energies using the fraction of the jet energy carried by each type of PFOs (blue histogram in fig \ref{fig:sigmaE}a) \cite{Thomson_2009,Ebrahimi:2017}. In presence of semi-leptonic decays, an additional uncertainty will be added by the neutrino correction ($\sigma_{\nu}$) \cite{Ebrahimi:2017}. The impact of the detector resolution for each PFO type and also the confusion in the particle flow has been studied in detail \cite{radkhorrami2021conceptual, Ebrahimi:2017}. The covariance matrix of the jet is then fed to kinematic fit as uncertainties on the measured the jet energy and angles. This new jet error parametrization improves the fit performance with a much better pull distribution of the jet energy (red histogram in fig \ref{fig:sigmaE}b) which receives further improvement (blue histogram in fig \ref{fig:sigmaE}b) by scaling up the jet energy error by a factor of 1.2 (derived from the width of the blue histogram in fig \ref{fig:sigmaE}a).
\section{$Z/H$ mass reconstruction}\label{SEC:ZHMassReco}
\begin{figure}[htbp] \centering
	\includegraphics[width=0.65\textwidth]{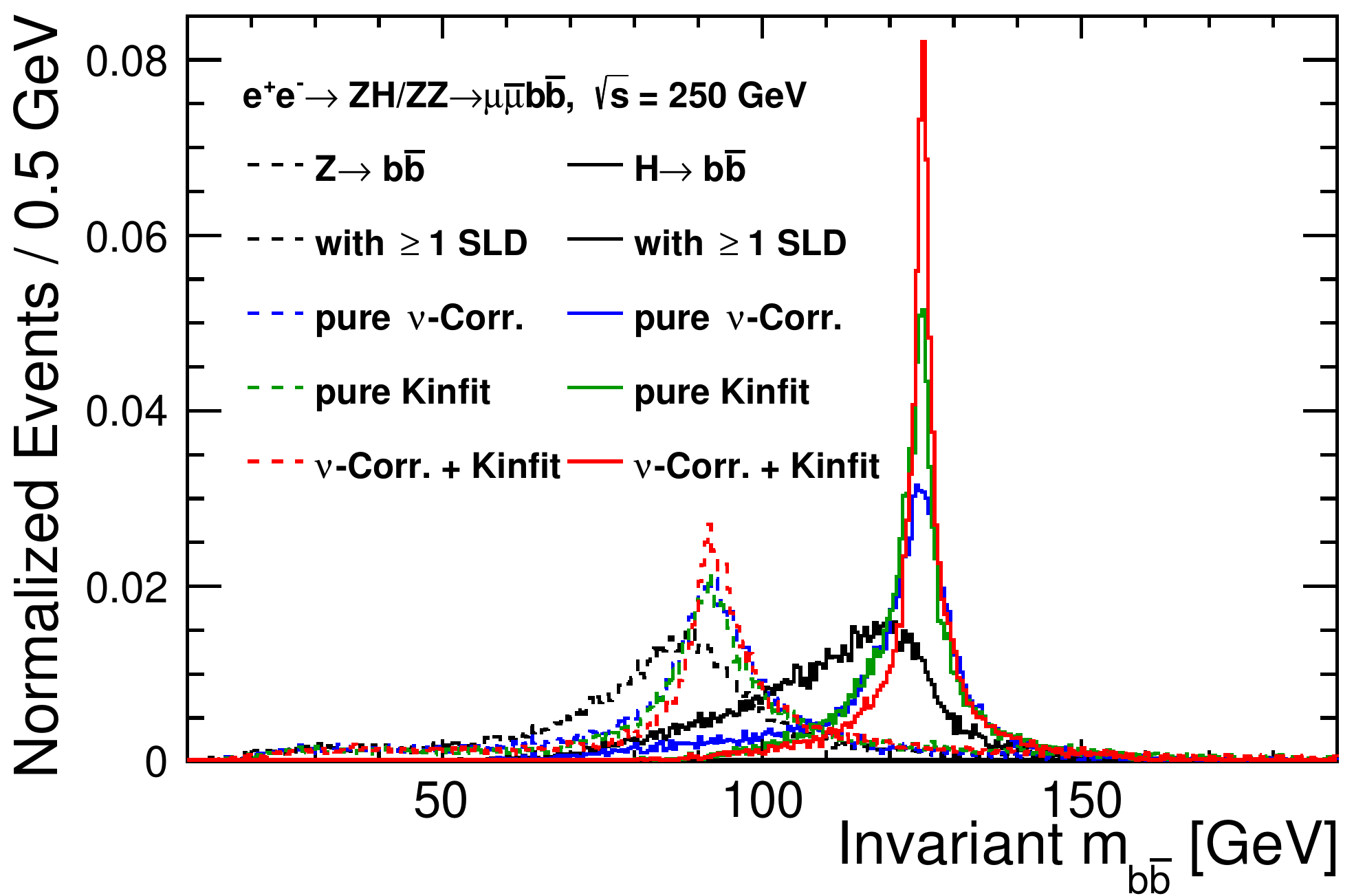}
	\caption{Invariant di-jet mass in $ZH$ and $ZZ$ events reconstructed by the PFA only (black), with neutrino correction (blue), with kinematic fit (green) and with neutrino correction and kinematic fit (red).}
	\label{fig:ZZHZ_InvMass}
\end{figure}

The new jet error parametrization and the neutrino correction, for the time being based on cheated inputs in terms of mass and decay point of the mother hadron, and association and momenta of the visible decay products, are applied on the introduced $ZH/ZZ\rightarrow\mu\bar{\mu}b\bar{b}$ samples. The blue histograms in fig \ref{fig:ZZHZ_InvMass} show the mass recovery by the neutrino correction alone for both Higgs boson (solid) and $Z$ boson (dashed). In this case, the kinematic fit is performed to resolve the sign ambiguity, but the pre-fit jet four-momenta are used for the invariant mass of di-jet. The green histograms in fig \ref{fig:ZZHZ_InvMass} show the mass distributions obtained from the kinematic fit only, without applying the neutrino correction. Even without the neutrino correction, the reconstructed di-jet masses receive a huge improvement from the new jet error parametrization described in section \ref{SEC:KinFit}. Finally, the red histograms show the combination of kinematic fit and (cheated) neutrino correction.
Even if the neutrino correction based on reconstructed information might be less powerful, the final performance is expected somewhere between the green and red histograms. In all cases, the $ZH$/$ZZ$ separation is improved drastically with respect to the black histograms, and the problem shown in fig \ref{fig:ZZHZ_Higgsmass} is avoided. This is important for any Higgs studies where event selection is done based on di-jet masses. The impact of the neutrino correction seems much smaller for $Z\rightarrow b\bar{b}$ than for $H\rightarrow b\bar{b}$. This is suspected to be related to the effect of the detector resolution (in addition to other sources of uncertainties) for the $H\rightarrow b\bar{b}$ distributions while the $Z\rightarrow b\bar{b}$ peak is limited by the natural width of the $Z$ boson.
\section{Conclusions}\label{SEC:Concl}
The separation of the $ZZH$ background and the $ZHH$ signal based on the invariant di-jet masses is very important for the Higgs self-coupling analysis. Since the Higgs decays predominently into $b\bar{b}$, the invariant di-jet mass is significantly affected by the reconstruction of heavy-flavor jets. A conceptual approach for a correction of semi-leptonic decays based on a detailed reconstruction of the decay kinematics can be combined with a kinematic fit. The propagation of PFO-level measurement uncertainties plus a parametrization of PFO confusion uncertainties, lead to a precise knowledge of the measurement errors on the jet level. The kinematic fit with the detailed error parametrization and the neutrino correction (so far using cheated inputs as proof-of-principle) have been tested on $ZH\rightarrow\mu\bar{\mu}b\bar{b}$ and $ZZ\rightarrow\mu\bar{\mu}b\bar{b}$ events.  Each of the techniques on its own and even more so their combination leads to a significant improvement of the $ZH$/$ZZ$ separation via the di-jet mass. In the future, the neutrino correction will be performed based on fully reconstructed information and both techniques will be applied to the Higgs self-coupling analysis.
\printbibliography[title=References]
\end{document}